\documentclass[runningheads]{llncs}
\usepackage[T1]{fontenc}
\usepackage{graphicx}
\usepackage{amsmath}
\usepackage{amssymb}
\begin{document}
\title{HyperAMS-Net: Adaptive Multi-Scale Spatial Hypergraph Network for Brain Disorder Classification}
\titlerunning{HyperAMS-Net for Brain Disorder Classification}
%
\author{
Proloy Kumar Mondal\inst{1}\orcidID{0009-0009-7991-4086} \and
Md Kamran Hussin Chowdhury\inst{2}\orcidID{0009-0008-1954-7600} \and
Hoi Leong Lee\inst{3}\thanks{Corresponding author.}\orcidID{0000-0002-4984-2183}
}


\authorrunning{P. K. Mondal et al.}

\institute{
Department of Computer Science and Engineering, University of North Texas, Denton, Texas, USA \and
Inje University, Gimhae-si, Republic of Korea \and
Faculty of Electronic Engineering and Technology, Universiti Malaysia Perlis, Arau, Malaysia \\
\email{hoileong@unimap.edu.my}
}

\maketitle
\begin{abstract}
Accurate classification of brain disorders from neuroimaging data remains challenging because of substantial inter-subject heterogeneity and the complex multi-scale patterns present in functional connectivity and morphological representations. To address these challenges, we propose HyperAMS-Net, a deep learning framework for brain disorder classification using neuroimaging representations derived from resting-state functional MRI or structural MRI. HyperAMS-Net integrates adaptive multi-scale convolution, hypergraph attention, spatial-channel attention, and adaptive feature fusion. Specifically, adaptive multi-scale convolution learns data-driven weights over multiple receptive fields to capture complementary patterns at different scales. Hypergraph attention models higher-order dependencies among learned feature representations through node--hyperedge--node message passing, while spatial-channel attention enhances discriminative feature learning. Adaptive feature fusion further aggregates complementary information across parallel network branches. HyperAMS-Net is evaluated on three benchmark datasets spanning distinct brain disorders: ABIDE for autism spectrum disorder, REST-meta-MDD for major depressive disorder, and ADNI for Alzheimer’s disease, using 5-fold stratified cross-validation. HyperAMS-Net achieves state-of-the-art performance across all evaluated datasets, attaining the highest accuracy and AUC among the compared methods. Ablation studies further demonstrate the contribution of each proposed component, with the largest performance degradation observed when hypergraph attention is removed.

\end{abstract}
\section{Introduction}

Neuroimaging has become increasingly important for investigating structural and functional alterations associated with brain disorders. Functional magnetic resonance imaging (fMRI) provides a non-invasive means of examining brain activity through blood-oxygen-level-dependent (BOLD) signals~\cite{ref1}. Altered functional connectivity patterns have been associated with neurodevelopmental and psychiatric disorders such as autism spectrum disorder (ASD) and major depressive disorder (MDD)~\cite{ref2,ref3}. These advances have motivated the development of computational methods for identifying disorder-related patterns from neuroimaging representations.

Conventional functional connectivity (FC) analysis commonly characterizes relationships between brain regions using pairwise measures such as Pearson or partial correlation coefficients~\cite{ref4,ref5}. More recently, graph-based deep learning methods have been increasingly applied to brain-network analysis, representing regions of interest (ROIs) as nodes and their relationships as edges~\cite{ref6,ref7}. Methods such as BrainGNN~\cite{ref8}, PopulationGCN~\cite{ref9}, and KMGCN~\cite{ref10} have demonstrated the effectiveness of graph-based representations for brain disorder classification. Noman et al.~\cite{ref11} further proposed AGMGC, which adaptively constructs brain graphs from BOLD time series using a learnable graph module.

Despite these advances, two challenges motivate the proposed framework. First, discriminative neuroimaging representations may contain informative patterns at different receptive-field scales, whereas conventional single-scale convolution may not effectively capture complementary patterns across multiple scales. Multi-scale feature architectures provide a potential means of addressing this limitation~\cite{ref12}. Second, conventional graph representations primarily encode pairwise relationships, whereas hypergraphs provide a mechanism for modeling higher-order interactions involving multiple nodes~\cite{ref13,ref14,ref15}. Incorporating such higher-order relational modeling with adaptive multi-scale feature extraction therefore provides a promising direction for brain disorder classification.


To address these challenges, we propose HyperAMS-Net, a hypergraph-guided adaptive multi-scale network for brain disorder classification. The framework comprises four components: (1) Adaptive Multi-Scale Convolution (AMSConv), which learns softmax-normalized scale weights across multiple receptive fields for adaptive multi-scale feature extraction; (2) Hypergraph Attention (HGA), which performs node–hyperedge–node message passing through learnable virtual hyperedges to capture higher-order feature dependencies; (3) Spatial-Channel Attention (SCA), which refines discriminative representations through channel and spatial attention; and (4) Adaptive Feature Fusion (AFF), which learns to aggregate complementary information across parallel network branches. We evaluate HyperAMS-Net on three benchmark neuroimaging datasets spanning ASD, MDD, and Alzheimer’s disease (AD), using FC representations for ABIDE and REST-meta-MDD datasets and structural MRI-derived morphological representations for ADNI dataset under 5-fold stratified cross-validation.

Our contributions are threefold: (1) we introduce AMSConv with learnable scale-importance weights for adaptive multi-scale feature extraction; (2) we develop HGA to model higher-order dependencies through learnable virtual hyperedges and node–hyperedge–node message passing; and (3) we integrate SCA and AFF into a unified multi-branch framework and conduct comprehensive evaluation and ablation analyses across three brain disorder benchmarks.

\section{Proposed Method}

\begin{figure}[t]
    \centering
    \includegraphics[width=1\columnwidth]{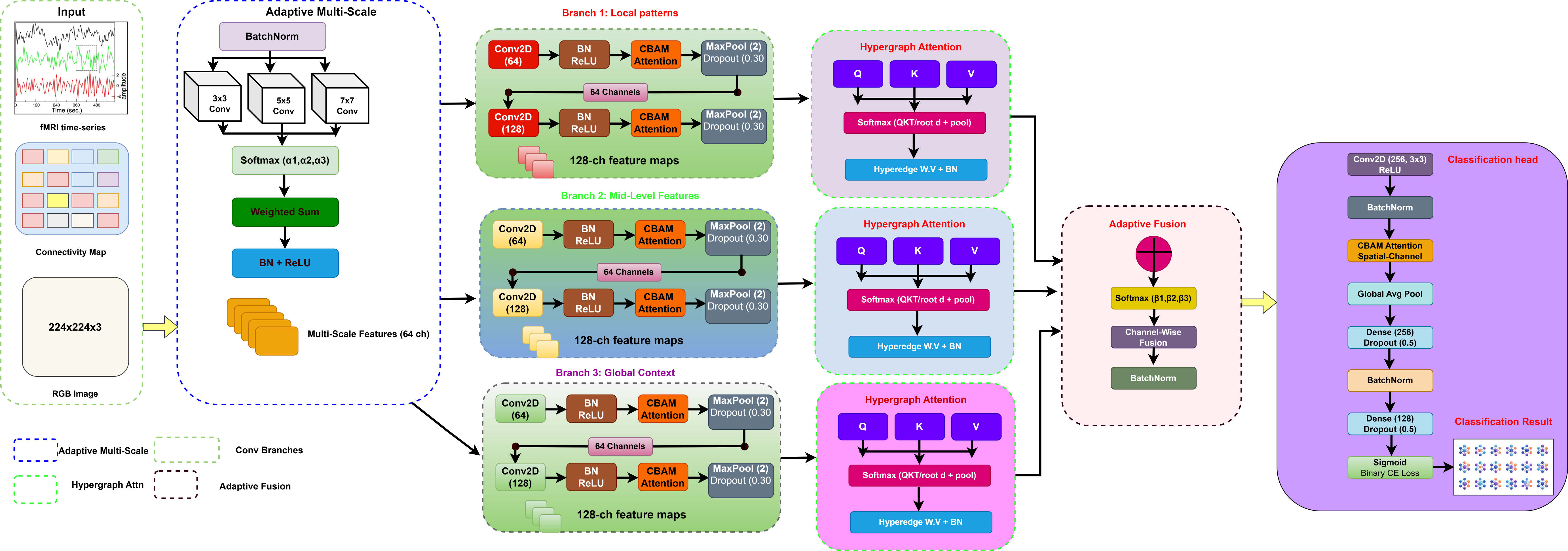}
    \caption{Overall architecture of HyperAMS-Net.}
    \label{fig:framework}
\end{figure}

We present HyperAMS-Net, a hypergraph-guided adaptive multi-scale network for brain disorder classification. As illustrated in Fig.~\ref{fig:framework}, the framework comprises four components: (1) Adaptive Multi-Scale Convolution (AMSConv), (2) Hypergraph Attention (HGA), (3) Spatial-Channel Attention (SCA), and (4) Adaptive Feature Fusion (AFF). The input is a FC matrix for ABIDE and REST-meta-MDD or an ROI morphological matrix for ADNI. Each input is represented as
$X \in \mathbb{R}^{B \times H \times W \times C}$
and resized to $224 \times 224$, with $C=3$ obtained through channel replication for compatibility with the convolutional backbone and CNN baselines. The ROI ordering is fixed within each dataset and kept consistent across subjects, ensuring correspondence between matrix locations. Consequently, convolutional neighborhoods capture local patterns within the ordered matrix representation and are not interpreted as direct anatomical proximity.

\subsection{Adaptive Multi-Scale Convolution (AMSConv)}

Neuroimaging connectivity and morphological representations may contain discriminative patterns at different receptive-field scales. Standard single-kernel convolutions may therefore be insufficient to capture both fine-grained and broader contextual patterns simultaneously. To address this limitation, we introduce AMSConv, which extracts features at $S=3$ receptive-field scales and combines them using learnable scale weights.

Given an input $\hat{X}$ after initial batch normalization, the feature map at scale $s$ is computed as
\begin{equation}
f_s =
\operatorname{ReLU}
\left(
\operatorname{BN}
\left(
W_s \ast \hat{X}
\right)
\right),
\qquad
s \in \{1,2,3\},
\qquad
k_s \in \{3,5,7\},
\label{eq:amsconv}
\end{equation}
where $W_s$ denotes a convolutional kernel of size $k_s \times k_s$, and $\ast$ denotes the convolution operation. Each branch produces $F=64$ feature maps. The input matrices are resized to $224 \times 224$ while preserving the dataset-specific ROI ordering. The outputs from the three scales are adaptively combined using softmax-normalized learnable weights:
\begin{equation}
Z_{\mathrm{AMS}}
=
\operatorname{BN}
\left(
\sum_{s=1}^{S}
\tilde{\alpha}_s f_s
\right),
\label{eq:ams_fusion}
\end{equation}
where
\begin{equation}
\tilde{\alpha}_s
=
\frac{\exp(\alpha_s)}
{\sum_{j=1}^{S}\exp(\alpha_j)}.
\label{eq:ams_weights}
\end{equation}

Let
\begin{equation}
\boldsymbol{\alpha}
=
[\alpha_1,\alpha_2,\alpha_3]
\in \mathbb{R}^{S},
\qquad S=3,
\label{eq:scale_logits}
\end{equation}
denote the learnable scale logits, with each $\alpha_s$ initialized to $1$ and optimized jointly with the network. Unlike sample-dependent gating mechanisms, the resulting scale weights are learned globally across the training data rather than being generated separately for each subject. This allows AMSConv to learn the relative importance of different receptive-field scales for the overall classification task. The resulting multi-scale representation $Z_{\mathrm{AMS}}$ is subsequently fed into the multi-branch network.

\subsection{Hypergraph Attention (HGA)}

Standard self-attention primarily models pairwise relationships between learned
feature tokens. To capture higher-order dependencies beyond pairwise interactions, we introduce a HGA block that performs node--hyperedge--node message passing in the learned feature space.

Let the input feature map to HGA be
$F_{\mathrm{in}} \in \mathbb{R}^{B \times h \times w \times C}$,
which is reshaped into a token sequence
$T \in \mathbb{R}^{B \times N \times C}$, where $N = h \times w$.
Query, key, and value projections are computed as
\begin{equation}
Q = TW_Q, \qquad
K = TW_K, \qquad
V = TW_V,
\end{equation}
where $W_Q$, $W_K$, and $W_V$ denote the query, key, and value projection
matrices, respectively. We define $M$ learnable virtual hyperedges. A node-to-hyperedge soft incidence
matrix is obtained through a learnable projection
$W_E \in \mathbb{R}^{C \times M}$ and normalized over the node dimension:
\begin{equation}
I =
\operatorname{softmax}_{N}(TW_E)
\in
\mathbb{R}^{B \times N \times M}.
\end{equation}

Node-to-hyperedge aggregation pools node features into hyperedge representations:
\begin{equation}
U =
I^{\top}V
\in
\mathbb{R}^{B \times M \times C}.
\end{equation}

A learnable hyperedge affinity matrix
$H \in \mathbb{R}^{M \times M}$
models interactions among the virtual hyperedges:
\begin{equation}
\tilde{U} =
\operatorname{softmax}(H)U
\in
\mathbb{R}^{B \times M \times C}.
\end{equation}

The refined hyperedge representations are then propagated back to the nodes and
combined with pairwise self-attention:
\begin{equation}
A =
\frac{QK^{\top}}{\sqrt{C}}
\in
\mathbb{R}^{B \times N \times N},
\qquad
T_{\mathrm{out}}
=
\operatorname{softmax}(A)V
+
I\tilde{U}.
\end{equation}

Here, $\operatorname{softmax}(A)V$ captures pairwise token dependencies, while
$I\tilde{U}$ injects higher-order group-wise context by projecting the
affinity-refined hyperedge representations back to each node according to its
incidence weights. The output is reshaped to
$B \times h \times w \times C$ and combined with the input through a residual
connection:
\begin{equation}
F_{\mathrm{HGA}}
=
\operatorname{BN}
\left(
\operatorname{reshape}(T_{\mathrm{out}})
+
F_{\mathrm{in}}
\right).
\end{equation}

In our implementation, $M=8$ learnable virtual hyperedges are used to provide
higher-order relational context complementary to the pairwise self-attention
pathway.


\subsection{Spatial-Channel Attention (SCA)}

We incorporate a dual-domain attention module inspired by the Convolutional
Block Attention Module (CBAM)~\cite{ref16}, applying sequential channel
and spatial attention to emphasize informative feature responses and suppress
irrelevant or noisy activations.

\textbf{Channel Attention.}
Given a feature map
$F \in \mathbb{R}^{H \times W \times C}$,
global average-pooled and max-pooled descriptors
$F_{\mathrm{avg}}$ and $F_{\mathrm{max}} \in \mathbb{R}^{C}$
are passed through a shared two-layer multilayer perceptron (MLP) with
reduction ratio $r=8$:
\begin{equation}
M_c(F)
=
\sigma
\left(
\operatorname{MLP}(F_{\mathrm{avg}})
+
\operatorname{MLP}(F_{\mathrm{max}})
\right)
\in
\mathbb{R}^{C}.
\end{equation}
The channel-refined feature map is $F' = F \otimes M_c(F)$, where
$\otimes$ denotes channel-wise multiplication.

\textbf{Spatial Attention.}
The channel-refined feature map is pooled along the channel dimension using
average and max pooling. The resulting descriptors are concatenated and
processed by a $7 \times 7$ convolution:
\begin{equation}
M_s(F')
=
\sigma
\left(
\operatorname{Conv}_{7 \times 7}
\left(
[
\operatorname{AvgPool}(F');
\operatorname{MaxPool}(F')
]
\right)
\right)
\in
\mathbb{R}^{H \times W \times 1}.
\end{equation}
The final spatial-channel refined representation is
$F_{\mathrm{SCA}} = F' \otimes M_s(F')$.

\subsection{Multi-Branch Architecture and Adaptive Feature Fusion (AFF)}

Following AMSConv, the multi-scale representation
$Z_{\mathrm{AMS}}$ is fed into three parallel branches, producing branch
feature maps
$b_i \in \mathbb{R}^{h \times w \times 128}$,
$i \in \{1,2,3\}$.
The branch outputs are fused through AFF:
\begin{equation}
F_{\mathrm{fused}}
=
\operatorname{BN}
\left(
\sum_{i=1}^{3}
\tilde{\phi}_i \odot b_i
\right),
\end{equation}
where the normalized fusion weights are given by
$\tilde{\phi}_i = \operatorname{softmax}_{i}(\Phi) \in \mathbb{R}^{128}$.

\subsection{Classification Head and Training Objective}

The fused representation is passed through the classification head, comprising
convolutional, pooling, and fully connected operations. Let
$F_{\mathrm{final}}$ denote the feature map provided to global average pooling
(GAP). The predicted probability is computed as
\begin{equation}
\hat{y}
=
\sigma
\left(
W_2 \cdot
\delta
\left(
\operatorname{BN}
\left(
W_1 \cdot \operatorname{GAP}(F_{\mathrm{final}}) + b_1
\right)
\right)
+
b_2
\right).
\end{equation}

The training objective is weighted binary cross-entropy:
\begin{equation}
\mathcal{L}
=
-\frac{1}{N_s}
\sum_{n=1}^{N_s}
\left[
\omega_1 y_n \log \hat{y}_n
+
\omega_0(1-y_n)\log(1-\hat{y}_n)
\right],
\end{equation}
where $N_s$ denotes the number of samples, and $\omega_0$ and $\omega_1$ are
inverse-frequency class weights computed from the training-fold distribution.

\section{Experiments and Results}
\subsection{Datasets}
We evaluate HyperAMS-Net on three benchmark neuroimaging datasets spanning ASD, MDD, and AD. The ABIDE dataset~\cite{ref17} contains resting-state fMRI data from 1,035 subjects collected across 17 international sites. FC matrices are constructed using the CC200 atlas with 200 ROIs, and the corresponding time series are truncated to $T=78$ after preprocessing. The REST-meta-MDD Consortium dataset~\cite{ref18} contains 1,601 resting-state fMRI scans acquired from 25 sites in China, with demographic and clinical heterogeneity across sites. The corresponding time series are truncated to $T=140$. For AD classification, we use data from the ADNI-1 cohort of the ADNI dataset~\cite{ref19}. Preprocessed T1-weighted structural MRI scans are used to extract region-level morphological features based on the Desikan--Killiany atlas with 84 ROIs. The task is formulated as binary classification between cognitively normal (CN) participants and patients with AD.

\subsection{Experimental Setup}

\textbf{Training and Evaluation.}
HyperAMS-Net is trained for a maximum of 100 epochs using AdamW with a learning
rate of $10^{-3}$ and weight decay of $5\times10^{-4}$. Gradient clipping is
applied with $\|\nabla\|\leq1.0$. We employ 5-fold stratified cross-validation,
with 15\% of each training fold reserved for validation. All preprocessing
statistics and hyperparameters are derived from the training data only. A ReduceLROnPlateau scheduler with a reduction factor of 0.5 and patience of
10 epochs is used. Early stopping is applied with a patience of 25 epochs based
on validation AUC. Models are trained using $224\times224$ inputs with a batch
size of 16. All baseline methods use identical data splits and preprocessing.
Performance is evaluated using accuracy (ACC), sensitivity (SEN),
specificity (SPE), precision (PRE), F1-score (F1), and area under the
receiver operating characteristic curve (AUC).

\begin{table}[t]
\centering
\caption{Performance comparison (mean $\pm$ std) on the ABIDE dataset.}
\label{tab:abide_comp}

\setlength{\tabcolsep}{6pt}        
\renewcommand{\arraystretch}{1.25} 

\resizebox{\textwidth}{!}{
\begin{tabular}{lcccccc}
\hline
\textbf{Method} 
& \textbf{ACC (\%)} & \textbf{PRE (\%)} & \textbf{SEN (\%)} 
& \textbf{SPE (\%)} & \textbf{F1 (\%)} & \textbf{AUC (\%)} \\
\hline
SVM 
& $67.34\pm2.26$ & $67.30\pm3.03$ & $70.78\pm2.70$ 
& $63.80\pm4.47$ & $68.94\pm1.87$ & $67.29\pm2.29$ \\

MLP 
& $65.12\pm6.13$ & $66.50\pm5.55$ & $64.43\pm8.20$ 
& $65.93\pm6.66$ & $65.33\pm6.36$ & $65.18\pm6.10$ \\

PopulationGCN~\cite{ref9}
& $80.73\pm3.04$ & $79.55\pm3.66$ & $85.59\pm5.52$ 
& $75.34\pm4.98$ & $82.35\pm3.09$ & $80.47\pm3.00$ \\

BrainGNN~\cite{ref8}
& $70.74\pm4.45$ & $71.91\pm4.95$ & $73.66\pm5.33$ 
& $67.54\pm7.76$ & $72.65\pm3.96$ & $70.60\pm4.53$ \\

PLSNet~\cite{ref20}
& $74.69\pm3.73$ & $74.96\pm1.57$ & $77.95\pm10.52$ 
& $70.87\pm5.41$ & $76.12\pm5.51$ & $83.88\pm3.46$ \\

KMGCN~\cite{ref10}
& $77.68\pm2.28$ & $80.99\pm7.13$ & $75.99\pm8.05$ 
& $79.44\pm11.37$ & $77.94\pm3.12$ & $85.88\pm4.82$ \\

GATE~\cite{ref21}
& $75.10\pm6.91$ & $74.71\pm4.99$ & $79.49\pm9.27$ 
& $70.21\pm4.87$ & $76.97\pm6.92$ & $74.85\pm6.78$ \\

AGMGC~\cite{ref11} 
& $84.65\pm3.36$ & $86.14\pm6.99$ & $85.43\pm6.15$ 
& $84.08\pm9.16$ & $85.44\pm2.86$ & $84.75\pm3.28$ \\

\textbf{HyperAMS-Net} 
& \textbf{86.82$\pm$2.14}
& \textbf{88.31$\pm$3.72}
& \textbf{87.65$\pm$4.08}
& \textbf{86.13$\pm$5.21}
& \textbf{87.73$\pm$2.31}
& \textbf{87.54$\pm$2.68} \\
\hline
\end{tabular}
}
\end{table}

\begin{table}[t]
\centering
\caption{Performance comparison (mean $\pm$ std) on the REST-meta-MDD dataset.}
\label{tab:mdd_comp}

\setlength{\tabcolsep}{6pt}
\renewcommand{\arraystretch}{1.25}

\resizebox{\textwidth}{!}{
\begin{tabular}{lcccccc}
\hline
\textbf{Method} 
& \textbf{ACC (\%)} & \textbf{PRE (\%)} & \textbf{SEN (\%)} 
& \textbf{SPE (\%)} & \textbf{F1 (\%)} & \textbf{AUC (\%)} \\
\hline
SVM 
& 64.71$\pm$1.59 & 64.70$\pm$1.47 & 70.36$\pm$3.02 
& 58.62$\pm$3.20 & 67.38$\pm$1.68 & 64.49$\pm$1.60 \\

MLP 
& 62.71$\pm$0.88 & 64.69$\pm$0.81 & 61.93$\pm$4.39 
& 63.56$\pm$3.38 & 63.20$\pm$2.05 & 62.74$\pm$0.79 \\

PopulationGCN~\cite{ref9}
& 72.30$\pm$1.55 & 72.66$\pm$1.37 & \textbf{89.34$\pm$1.90} 
& 43.93$\pm$4.12 & 80.12$\pm$1.07 & 66.63$\pm$1.92 \\

BrainGNN~\cite{ref8} 
& 63.63$\pm$1.16 & 66.60$\pm$2.63 & 84.84$\pm$8.00 
& 28.23$\pm$13.97 & 74.35$\pm$2.05 & 56.53$\pm$3.19 \\

PLSNet~\cite{ref20} 
& 71.25$\pm$1.16 & 75.37$\pm$3.86 & 82.34$\pm$7.75 
& 51.87$\pm$13.48 & 78.39$\pm$1.79 & 75.94$\pm$1.15 \\

KMGCN~\cite{ref10} 
& 68.75$\pm$5.28 & 74.65$\pm$3.65 & 77.12$\pm$7.51 
& 54.05$\pm$7.75 & 75.74$\pm$4.67 & 72.19$\pm$5.98 \\

GATE~\cite{ref21} 
& 73.79$\pm$1.43 & 78.30$\pm$1.59 & 80.37$\pm$0.45 
& 62.85$\pm$3.32 & 79.31$\pm$0.96 & 71.61$\pm$1.80 \\

AGMGC~\cite{ref11} 
& 82.36$\pm$2.84 & 82.11$\pm$3.20 & 80.94$\pm$6.11 
& 83.64$\pm$3.90 & 81.40$\pm$3.44 & 82.29$\pm$2.93 \\

\textbf{HyperAMS-Net} 
& \textbf{84.17$\pm$2.03} & \textbf{83.85$\pm$2.76} & 83.62$\pm$3.94
& \textbf{84.73$\pm$3.15} & \textbf{83.52$\pm$2.18} & \textbf{84.31$\pm$2.47} \\
\hline
\end{tabular}
}
\end{table}

\begin{table}[t]
\centering
\caption{Performance comparison (mean $\pm$ std) on the ADNI dataset.}
\label{tab:adni_comp}

\setlength{\tabcolsep}{6pt}
\renewcommand{\arraystretch}{1.25}

\resizebox{\textwidth}{!}{
\begin{tabular}{lcccccc}
\hline
\textbf{Method} 
& \textbf{ACC (\%)} & \textbf{PRE (\%)} & \textbf{SEN (\%)} 
& \textbf{SPE (\%)} & \textbf{F1 (\%)} & \textbf{AUC (\%)} \\
\hline
SVM 
& 71.24$\pm$3.15 & 70.88$\pm$3.47 & 73.52$\pm$4.21 
& 69.04$\pm$5.32 & 72.12$\pm$3.58 & 71.18$\pm$3.09 \\

MLP 
& 69.47$\pm$4.82 & 70.13$\pm$5.11 & 68.79$\pm$6.34 
& 70.26$\pm$5.88 & 69.38$\pm$5.07 & 69.52$\pm$4.76 \\

ResNet-50~\cite{ref22}  
& 74.83$\pm$3.26 & 74.21$\pm$3.78 & 76.35$\pm$4.13 
& 73.42$\pm$4.61 & 75.26$\pm$3.55 & 75.14$\pm$3.41 \\

EfficientNetB0~\cite{ref23}  
& 76.51$\pm$2.94 & 75.87$\pm$3.42 & 78.13$\pm$3.87 
& 74.96$\pm$4.23 & 76.97$\pm$3.12 & 77.08$\pm$2.88 \\

BrainGNN~\cite{ref8} 
& 73.16$\pm$4.07 & 74.05$\pm$4.53 & 74.88$\pm$5.14 
& 71.43$\pm$6.22 & 74.46$\pm$4.31 & 73.24$\pm$4.12 \\

GATE~\cite{ref21} 
& 77.34$\pm$3.51 & 76.92$\pm$3.84 & 79.61$\pm$4.07 
& 75.12$\pm$4.87 & 78.23$\pm$3.62 & 77.89$\pm$3.44 \\

KMGCN~\cite{ref10}  
& 78.92$\pm$2.73 & 79.45$\pm$3.16 & 80.37$\pm$3.64 
& 77.56$\pm$4.18 & 79.90$\pm$2.91 & 79.48$\pm$2.65 \\

AGMGC~\cite{ref11} 
& 81.64$\pm$2.31 & 82.17$\pm$2.87 & 83.08$\pm$3.25 
& 80.27$\pm$3.74 & 82.62$\pm$2.47 & 82.19$\pm$2.38 \\

\textbf{HyperAMS-Net} 
& \textbf{84.73$\pm$1.97} & \textbf{85.32$\pm$2.41} & \textbf{85.86$\pm$2.93} 
& \textbf{83.68$\pm$3.27} & \textbf{85.59$\pm$2.14} & \textbf{85.12$\pm$1.88} \\
\hline
\end{tabular}
}
\end{table}

\begin{table}[!t]
\centering
\caption{Statistical significance analysis (paired $t$-test) between HyperAMS-Net and competing methods.}
\label{tab:t_test}
\resizebox{\textwidth}{!}{%
\begin{tabular}{lcccccc}
\hline
\textbf{Comparison} 
& \textbf{ABIDE $t$} & \textbf{ABIDE $p$} 
& \textbf{MDD $t$} & \textbf{MDD $p$} 
& \textbf{ADNI $t$} & \textbf{ADNI $p$} \\
\hline
vs SVM            & 8.42 & \textbf{$<$0.001} & 7.83 & \textbf{$<$0.001} & 9.14 & \textbf{$<$0.001} \\
vs MLP            & 9.17 & \textbf{$<$0.001} & 8.65 & \textbf{$<$0.001} & 9.88 & \textbf{$<$0.001} \\
vs PopulationGCN~\cite{ref9}  & 4.23 & \textbf{0.013} & 5.31 & \textbf{0.006} & 6.12 & \textbf{0.004} \\
vs BrainGNN~\cite{ref8}      & 6.84 & \textbf{$<$0.001} & 7.12 & \textbf{$<$0.001} & 7.53 & \textbf{$<$0.001} \\
vs PLSNet~\cite{ref20}        & 5.17 & \textbf{0.007} & 4.89 & \textbf{0.009} & 5.64 & \textbf{0.005} \\
vs KMGCN~\cite{ref10}         & 4.56 & \textbf{0.010} & 4.13 & \textbf{0.015} & 4.78 & \textbf{0.008} \\
vs GATE~\cite{ref21}          & 5.34 & \textbf{0.006} & 4.67 & \textbf{0.011} & 5.21 & \textbf{0.007} \\
vs AGMGC~\cite{ref11}         & 3.14 & \textbf{0.035} & 3.27 & \textbf{0.031} & 3.48 & \textbf{0.028} \\
\hline
\end{tabular}%
}
\end{table}

\begin{table}[t]
\centering
\caption{Ablation study results (mean $\pm$ std) on ABIDE, REST-meta-MDD, and ADNI datasets.}
\label{tab:ablation}
\resizebox{\textwidth}{!}{
\begin{tabular}{lcccccc}
\hline
\textbf{Configuration} 
& \textbf{ABIDE ACC} & \textbf{ABIDE AUC} 
& \textbf{MDD ACC} & \textbf{MDD AUC} 
& \textbf{ADNI ACC} & \textbf{ADNI AUC} \\
\hline
\textbf{Full HyperAMS-Net} 
& \textbf{$86.82\pm2.14$} & \textbf{$87.54\pm2.68$} 
& \textbf{$84.17\pm2.03$} & \textbf{$84.31\pm2.47$} 
& \textbf{$84.73\pm1.97$} & \textbf{$85.12\pm1.88$} \\

w/o AMSConv 
& $83.47\pm2.61$ & $84.23\pm2.95$ 
& $81.35\pm2.44$ & $81.78\pm2.83$ 
& $81.96\pm2.38$ & $82.47\pm2.21$ \\

w/o HGA
& $82.91\pm2.83$ & $83.68\pm3.12$ 
& $80.74\pm2.67$ & $81.12\pm3.04$ 
& $81.28\pm2.57$ & $81.89\pm2.43$ \\

w/o SCA 
& $84.15\pm2.47$ & $85.01\pm2.73$ 
& $82.43\pm2.28$ & $82.87\pm2.64$ 
& $82.91\pm2.19$ & $83.34\pm2.07$ \\

w/o AFF
& $83.64\pm2.72$ & $84.46\pm2.97$ 
& $81.88\pm2.53$ & $82.19\pm2.78$ 
& $82.37\pm2.31$ & $82.94\pm2.16$ \\

CNN Backbone Only
& $79.32\pm3.44$ & $80.17\pm3.67$ 
& $77.56\pm3.18$ & $78.23\pm3.41$ 
& $78.64\pm3.02$ & $79.35\pm3.24$ \\
\hline
\end{tabular}
}
\end{table}

\subsection{Analysis of results} 

Table~\ref{tab:abide_comp} compares HyperAMS-Net with eight baseline methods on the ABIDE dataset. HyperAMS-Net achieves the best performance across all six evaluation metrics, with an ACC of 86.82\% and an AUC of 87.54\%. Its ACC exceeds AGMGC, the strongest competing method in terms of accuracy, by 2.17 percentage points. For AUC, HyperAMS-Net exceeds the strongest competing result, KMGCN (85.88\%), by 1.66 percentage points.

Table~\ref{tab:mdd_comp} summarizes the results on REST-meta-MDD. HyperAMS-Net achieves the highest ACC, PRE, SPE, F1, and AUC, with an ACC of 84.17\% and an AUC of 84.31\%. Although PopulationGCN and BrainGNN achieve higher sensitivity,
their specificity is substantially lower, whereas HyperAMS-Net maintains a
more balanced performance between SEN and SPE. Compared with AGMGC,
HyperAMS-Net improves ACC by 1.81 percentage points and AUC by 2.02
percentage points. Table~\ref{tab:adni_comp} presents the results on ADNI. HyperAMS-Net achieves the highest performance across all six evaluation metrics, attaining an ACC of 84.73\% and an AUC of 85.12\%. Compared with AGMGC, the strongest competing method in Table~3, HyperAMS-Net improves ACC by 3.09 percentage points and AUC by 2.93 percentage points.

Table~\ref{tab:t_test} reports the paired $t$-test comparisons between HyperAMS-Net and the competing methods. The reported results show $p<0.05$ for all listed comparisons across ABIDE, REST-meta-MDD, and ADNI, providing statistical support for the observed performance differences. Finally, Table~\ref{tab:ablation} presents the ablation study across the three datasets. The full HyperAMS-Net achieves the highest ACC and AUC on ABIDE,
REST-meta-MDD, and ADNI. Among the individual component ablations, removing
HGA produces the largest performance degradation across all three datasets, while the CNN-only backbone yields the lowest overall performance. These results indicate that AMSConv, HGA, SCA, and AFF each contribute to the final classification performance.



\section{Conclusion}

In this paper, we proposed HyperAMS-Net, a framework integrating AMSConv,
HGA, SCA, and AFF for brain disorder classification from neuroimaging data.
Experiments on three benchmark datasets, ABIDE, REST-meta-MDD, and ADNI,
show that HyperAMS-Net achieves the highest ACC and AUC among the compared
methods across all three datasets. The reported paired $t$-test results further
provide statistical support for the observed performance differences. The ablation study indicates that each proposed component contributes to the
overall performance, with the largest degradation among the individual
component removals observed when HGA is removed. These findings suggest that
combining adaptive multi-scale feature extraction, higher-order relational
modeling, SCA, and AFF is effective for brain disorder classification. The current evaluation is based on subject-level stratified cross-validation.
Future work should therefore validate generalization using site-stratified or
leave-one-site-out protocols to better isolate acquisition-site effects.

\begin{credits}
\subsubsection{\ackname} This work was supported in part by the Ministry of Higher Education Malaysia through the Fundamental Research Grant Scheme (FRGS) under Grant No. FRGS/1/2025/ICT02/UNIMAP/02/2.

\subsubsection{\discintname}
The authors have no competing interests to declare that are relevant to the content of this article.
\end{credits}






\bibliographystyle{splncs04}
\bibliography{MlMI_78}
\end{document}